\documentclass[a4paper,11pt]{article}

\newenvironment{proof}{\paragraph{Proof:}}{\hfill$\square$}
\newcommand{\maximize}{\mathrm{maximize}}

\newcommand{\subjectto}{\mathrm{subject~to}}

\newtheorem{prop}{Proposition}
\usepackage{url}
\usepackage{amsmath,amssymb}
\usepackage{bm}
\usepackage{graphicx}
\usepackage{authblk}
\usepackage{natbib}
\setcitestyle{authoryear,open={(},close={)}}
\usepackage[top=3cm,bottom=3cm,left=2.5cm,right=2.5cm]{geometry}

\title{An efficient branch-and-cut algorithm\\ for approximately submodular function maximization}

\author[1,2]{Naoya Uematsu \footnote{\texttt{naoya.uematsu@riken.jp}}}
\author[1,2]{Shunji Umetani \footnote{\texttt{umetani@ist.osaka-u.ac.jp}}}
\author[1,3]{Yoshinobu Kawahara \footnote{\texttt{kawahara@imi.kyushu-u.ac.jp}}}
\affil[1]{RIKEN Center for Advanced Intelligence Project, Quantative Biology Center, 6-2-4 Fruedai, Suita, Osaka, 565-0874, Japan.}
\affil[2]{Graduate School of Information Science and Technology, Osaka University, 1-5 Yamadaoka, Suita, Osaka, 565-0871, Japan.}
\affil[3]{Institute of Mathematics for Industry, Kyushu University, 744 Motooka, Fukuoka, Fukuoka, 819-0395, Japan.}

\date{}

\begin{document}
\maketitle

\begin{abstract}
When approaching to problems in computer science, we often encounter situations where a subset of a finite set maximizing some utility function needs to be selected. Some of such utility functions are known to be approximately submodular.
For the problem of maximizing an approximately submodular function (ASFM problem), a greedy algorithm quickly finds good feasible solutions for many instances while guaranteeing ($1-e^{-\gamma}$)-approximation ratio for a given submodular ratio $\gamma$.
However, we still encounter its applications that ask more accurate or exactly optimal solutions within a reasonable computation time.
In this paper, we present an efficient branch-and-cut algorithm for the non-decreasing ASFM problem based on its binary integer programming (BIP) formulation with an exponential number of constraints.
To this end, we first derive a BIP formulation of the ASFM problem and then, develop an improved constraint generation algorithm that starts from a reduced BIP problem with a small subset of constraints and repeats solving the reduced BIP problem while adding a promising set of constraints at each iteration.
Moreover, we incorporate it into a branch-and-cut algorithm to attain good upper bounds while solving a smaller number of nodes of a search tree.
The computational results for three types of well-known benchmark instances shows that our algorithm performs better than the conventional exact algorithms.
\end{abstract}

\section{INTRODUCTION}
When approaching to problems in computer science, we often encounter situations where a subset of a finite set maximizing some utility function needs to be selected.
Some of such utility functions are known to be submodular (e.g., sensor placement~\citep{Golovin2011,Kawahara2009,Kratica2001}, document summarization~\citep{Lin2011}, and influence spread problems~\citep{Kempe2003,Sakaue2018}).
A set function $f\colon 2^N\to \mathbb{R}$ is called submodular if it satisfies $ f(S \cup \{i\}) - f(S) \geq f(T \cup \{i\}) - f(T)$ for all $S \subseteq T \subseteq N $ and $ i \notin T$, where $N := \{ 1,\dots,n \}$ is a finite set.
Submodular functions can be considered as discrete counterparts of convex functions through the continuous relaxation called the Lov\'{a}sz extension \citep{Lovasz1983}.


Meanwhile, in many practical situations, utility functions may not necessarily be submodular. 
However even in those cases, submodularity can be approximately satisfied in various problems such as feature selection~\citep{Das2011,Yu2004}, boosting influence spread~\citep{Lin2017}, data summarization~\citep{Balkanski2016} and combinatorial auction~\citep{Conitzer2005}.
For this reason, the optimization of {\em an approximately submodular function} has been attracted an increasing attention recently~\citep{Das2018, Horel2016, Krause2014}.
This type of function is defined with {\em a submodular ratio} $\gamma$, which defined for a set function $f$ as the maximum value $0 < \gamma \le 1$ such that $ f(S \cup \{i\}) - f(S) \geq \gamma~(f(T \cup \{i\}) - f(T))$, for all $S \subseteq T \subseteq N $ and $  i \notin T$. 
That is, a submodular ratio $\gamma$ measures how close the function is to submodular~\citep{Das2011,Johnson2016}.

In this paper, we address the problem of maximizing a non-decreasing approximately submodular function $f$ under a cardinality constraint (hereafter, referred to as approximately submodular function maximization (ASFM) problem):
\begin{equation}
\label{eq:asfm}
\begin{array}{ll}
\maximize & f(S) \\
\subjectto &  |S| \leq k, \;\; S \subseteq N,  \\
\end{array}
\end{equation}
where $k \le n$ is a positive integer comprising the cardinality constraint.
A set function is non-decreasing if $f(S) \leq f(T)$ for all $S \subseteq T$ and $f(\emptyset) = 0$. 
\cite{Das2011} presented a greedy algorithm for the ASFM problem that guarantees $(1-e^{-\gamma})$-approximation ratio for a given submodular ratio $\gamma$ .
\cite{Chen2015} proposed an $\textnormal{A}^{\ast}$ search algorithm to obtain an exactly optimal solution for the ASFM problem. 
Their algorithm computes an upper bound by a variant of variable fixing techniques with $O(n)$ oracle queries.
Their algorithm quickly finds upper bounds; however the attained upper bounds are not often tight enough to prune nodes of the search tree effectively. 
Therefore, their algorithm often processes a huge number of nodes of the search tree until obtaining an optimal solution.

Here, we present an efficient branch-and-cut algorithm for the ASFM problem based on its binary integer programming (BIP) formulation with an exponential number of constraints.
To this end, we first derive a BIP formulation of the ASFM problem and then, develop a modified constraint generation algorithm based on the BIP formulation.
Unfortunately, the modified constraint generation algorithm is not efficient because of a large number of reduced BIP problems to be solved.
To overcome this, we propose an improved constraint generation algorithm, where a promising set of constraints is added at each iteration.
We further incorporate it into a branch-and-cut algorithm to attain good upper bounds while solving a smaller number of reduced BIP problems. 
Finally, we evaluate our algorithms under comparisons with the existing ones using three types of well-known benchmark instances and the combinatorial auction problem.

The remainder of this paper is organized as follows. 
First, in Section~\ref{sec:existing}, we give a brief review of the existing algorithms. In~Section~\ref{sec:IP_form}, we derive the IP formulation of the ASFM problem. Then, in Section~\ref{sec:proposed}, we propose three algorithms for solving the ASFM problem. We illustrate the effectivity of the proposed algorithm with the combinatorial auction problem in Section~\ref{sec:example} and show some computational results using three types of well-known benchmark instances in Section~\ref{sec:results}. Finally, the paper is concluded in Section~\ref{sec:concl}.

\section{EXISTING ALGORITHMS\label{sec:existing}}
Here, we first review the constraint generation algorithm by \citeauthor{Nemhauser1981} (\citeyear{Nemhauser1981}) for the problem~\eqref{eq:asfm} when $f$ is not approximately but {\em exactly} submodular (referred to as submodular function maximization (SFM) problem) in Subsection~\ref{ssec:cg} and then, $\textnormal{A}^\ast$~search algorithm proposed by \citeauthor{Chen2015} (\citeyear{Chen2015}) for the ASFM problem in Subsection~\ref{ssec:a-search}.

\subsection{Constraint Generation Algorithm for the SFM Problem}
\label{ssec:cg}

 \cite{Nemhauser1981} have proposed an exact algorithm for the SFM problem, called the constraint generation algorithm. The algorithm starts from a reduced BIP problem with a small subset of constraints and then, repeats solving the reduced BIP problem while adding a new constraint at each iteration.

Given a set of feasible solutions $Q \subseteq F$, we define $\textnormal{BIP}(Q)$ as the following reduced BIP problem of the SFM problem:
\begin{equation}
\label{eq:BIPQ}
\begin{array}{lll}
\textnormal{maximize} & z \\
\textnormal{subject to} & z  \leq  f(S) + \displaystyle\sum_{i \in N \setminus S} f(\{ i \} \mid S)  \; y_{i}, \; S \in Q,\\
& \displaystyle\sum_{i \in N} y_{i} \leq k,  \\
&  y_{i} \in \{0, 1\}, \; i \in N.
\end{array}
\end{equation}
The initial solution $S^{(0)}$ is obtained by applying the greedy algorithm \citep{Minoux1978,Nemhauser1978}.
Their algorithm starts with a set $Q=  \{S_{[0]}^{(0)},\dots, S_{[k]}^{(0)}\}$, where  $S_{[i]}$ denotes the first $i$ elements of a feasible solution $S^{(0)}$ with the order obtained by the greedy algorithm.
We now consider the $t$-th iteration of the constraint generation algorithm.
The algorithm first solves $\textnormal{BIP}(Q)$ with $Q = \{S_{[0]}^{(0)},\dots, S_{[k-1]}^{(0)}, S^{(0)},\dots,S^{(t-1)}\}$ to obtain an optimal solution $\boldsymbol{y}^{(t)} = (y_1^{(t)},\dots,y_n^{(t)})$ and the optimal value $z^{(t)}$ that gives an upper bound of that of the problem (\ref{eq:BIPQ}).
Let $S^{(t)}$ denote the optimal solution of $\textnormal{BIP}(Q)$ corresponding to $\boldsymbol{y}^{(t)}$, and $S^{\ast}$ denote the incumbent solution of the problem (\ref{eq:BIPQ}) obtained so far.
If $f(S^{(t)}) > f(S^{\ast})$ holds, then the algorithm replaces the incumbent solution  $S^{\ast}$ with $S^{(t)}$. 
If $z^{(t)} > f(S^{(t)})$ holds, the algorithm concludes $S^{(t)} \notin Q$ and adds $S^{(t)}$ to $Q$, because $S^{(t)}$ does not satisfy any constraints of $\textnormal{BIP}(Q)$.
That is, the algorithm adds the following constraint to $\textnormal{BIP}(Q)$ for improving the upper bound $z^{(t)}$ of the optimal value of the problem (\ref{eq:BIPQ}).
\begin{equation}
z \le f(S^{(t)}) + \sum_{i \in N \setminus S^{(t)}} f(\{ i \} \mid S^{(t)}) \; y_i.
\end{equation}
%
These procedures are repeated until $z^{(t)}$ and $f(S^{\ast})$ meet.

The pseudo code of this algorithm is shown below. We note that the value of $z^{(t)}$ is non-increasing with the number of iterations and the algorithm must terminate after at most $\binom{n}{k}$ iterations.
\smallskip
\begin{description}
\item[\underline{\textbf{Algorithm CG}$(S^{(0)})$}]
\item[\textbf{Input:}]~ The initial feasible solution $S^{(0)}$. 
\item[\textbf{Output:}]~ The incumbent solution $S^{\ast}$.
\item[\textbf{Step1:}]~ Set $Q \leftarrow \{ S_{[0]}^{(0)},\dots, S_{[k]}^{(0)} \}$, $S^{\ast} \leftarrow S^{(0)}$ and $t \leftarrow 1$. 
\item[\textbf{Step2:}]~ Solve $\textnormal{BIP}(Q)$.
Let $S^{(t)}$ and  $z^{(t)}$ be an optimal solution and the optimal value of $\textnormal{BIP}(Q)$, respectively.
\item[\textbf{Step3:}]~ If $ f(S^{(t)}) > f(S^{\ast})$ holds, then set $S^\ast  \leftarrow S^{(t)}$.
\item[\textbf{Step4:}]~ If $z^{(t)} = f(S^{\ast})$ holds, then output the incumbnet solution $S^{\ast} $ and exit. Otherwise; (i.e., $z^{(t)} > f(S^{\ast}) \ge f(S^{(t)})$), set $Q \leftarrow Q \cup \{ S^{(t)}\}$, $t \leftarrow t +1$ and return to Step2.
\end{description}

\subsection{$\textnormal{A}^{\ast}$~Search Algorithm for the ASFM Problem}
\label{ssec:a-search}

\cite{Chen2015} have proposed an $\textnormal{A}^{\ast}$ search algosithm for the ASFM problem.
We first define the search tree of the $\textnormal{A}^{\ast}$~search algorithm.
Each node $S$ of the search tree represents a feasible solution, where the root node is set to $S \leftarrow \emptyset$.
The parent of a node $T$ is defined as $S = T \setminus \{ T_{\max} \}$, where $T_{\max}$ is an element $i \in T$ with the largest number.
For example, node $S = \{ 3 \}$ is the parent of node $T = \{ 3, 5 \}$, since $T \setminus \{ T_{\max} \}= \{ 3,5 \} \setminus \{ 5 \} = \{ 3 \} = S$. 
The $\textnormal{A}^{\ast}$~search algorithm employs a list $L$ to manage nodes of the search tree.
The value of a node $S$ is defined as $\bar{f}(S) = f(S) + h(S)$, where $h(\cdot)$ is a heuristic function.
We note that $\bar{f}(\cdot)$ give an upper bound of the optimal value of the SFM problem at the node $S$.

The initial feasible solution is obtained by the greedy algorithm \citep{Minoux1978,Nemhauser1978}.
The algorithm repeats to extract a node $S$ with the largest value $\bar{f}(\cdot)$ from the list $L$ and insert its children $T \in F$ into the list $L$ at each iteration.
Let $S \in F$ be a node extracted from the list $L$, and $S^{\ast}$ be the incumbent solution (i.e., best feasible solution obtained so far).
The algorithm obtains a feasible solution $S^{\prime} \in F$ from the node $S$, e.g. a variety of greedy algorithms.
If $f(S^{\prime}) > f(S^{\ast})$ holds, then the algorithm replaces the incumbent solution $S^{\ast}$ with $S^{\prime}$.
Then, all children $T \in F$ of the node $S$ satisfying $\bar{f}(T) > f(S^{\ast})$ are inserted into the list $L$.
The algorithm repeats these procedures until the list $L$ becomes empty.

The pseudo code of this algorithm is shown below.
\smallskip
\begin{description}
\item[\underline{\textbf{Algorithm A$^{\ast}$}$(S)$}] 
\item[\textbf{Input:}]~ The initial feasible solution $S$.
\item[\textbf{Output:}]~ The incumbent solution $S^{\ast}$.
\item[\textbf{Step1:}]~ Set $L \leftarrow \{ \emptyset \}$ and $S^{\ast} \leftarrow S$.
\item[\textbf{Step2:}]~ If $L = \emptyset$ holds, then output the incumbent solution $S^{\ast}$ and exit. 
\item[\textbf{Step3:}]~ Extract a node $S$ with the largest value $\bar{f}(\cdot)$ from the list $L$.
If $\bar{f}(S) \leq f(S^{\ast})$ holds, then return to Step~2.
\item[\textbf{Step4:}]~ Obtain a feasible solution $S^{\prime} \in F$ from the node $S$.
If $f(S^{\prime}) > f(S^{\ast})$ holds, then set $S^{\ast} \leftarrow S^{\prime}$.
\item[\textbf{Step5:}]~ Set $L \leftarrow L \cup \{ T \}$ for all children $T$ of the node $S$ satisfying $T \in F$ and $\bar{f}(T) > f(S^{\ast})$.
Return to Step2.
\end{description}
\smallskip

We then illustrate a heuristic function $h(\cdot)$ applied to the $\textnormal{A}^{\ast}$~search algorithm.
Let $S$ be the current node of the $\textnormal{A}^{\ast}$~search algorithm. 
We consider the following reduced problem of the SFM problem for obtaining $h(\cdot)$.
\begin{equation}
\label{eq:red}
\begin{array}{ll}
\maximize & f_{S}(T) \\
\subjectto & T \subseteq N \setminus S^+, |T| \leq k - |S|,
\end{array}
\end{equation}
where $S^+ = \{i \in N \mid i \leq S_{\max} \}$ and $f_S(\cdot) = f(\cdot \mid S)$.
Let $T^\ast$ be an optimal solution of the reduced problem (\ref{eq:red}). 
By approximately submodularity, we obtain $\sum_{i \in T} (1 / \gamma)  f_S(\{ i \}) \geq f_{S}(T)$ for any $T \subseteq N$ and the following inequality. 
\begin{equation}
\max_{T \subseteq N \setminus S^+, |T| \leq k - |S| } \frac{1}{\gamma} \displaystyle\sum_{i \in T} f_{S}(\{ i \}) \geq  \frac{1}{\gamma} \sum_{i \in T^\ast} f_{S}( \{ i\} )  \geq  f_{S}(T^\ast).
\label{eq:VF}
\end{equation}
Since the reduced problem (\ref{eq:red}) is still NP-hard, we consider obtaining an upper bound of $f_{S}(T^\ast)$.
Let $\bar{S}^+$ be the non-increasing ordered set with respect to $f_{S}(\{ i \})$ for $i \in N \setminus S^+$.
We assume that $|S \cup \bar{S}^+| > k$, because we can obtain the upper bound by computing $f(S \cup \bar{S}^+)$ in otherwise.
Let $[p] = \{ 1,\dots, p\}$ and $\bar{S}^+_{[p]}$ denote the set of the first $p = k - |S|$ elements of the sorted set $\bar{S}^+$. 
We then define a heuristic function $h(\cdot)$ by
\begin{equation}
h(S) = \frac{1}{\gamma} \sum_{i \in \bar{S}^+_{[p]}} f_{S}(\{ i \}).
\end{equation}
We note that we let $\bar{S}^+_{[p]} \cup S$ be a feasible solution $S^{\prime} \in F$ for the node $S$ (Step 4).
If $f_S(\{ i \}) = 0$ holds for some $i \in \bar{S}^+_{[p]}$, then we conclude $f_{S}(\bar{S}^+_{[p]}) = f_{S}(T^\ast)$ by submodularity.
For a given node $S$, we compute an upper bound $\bar{f}(S) = f(S) + h(S)$.

%
%
\section{IP FORMULATION \label{sec:IP_form} }
In this section, we formulate the ASFM problem into a BIP problem.
First, the submodular ratio $\gamma$ is obtained as follows:
\begin{equation}
\gamma = \min_{S, T \subseteq N}  \frac{f(S) - f(S \cap T)}{f(S \cup T ) -f(T)},
\end{equation}
where we regard $0/0 = 1$.
According to \cite{Johnson2016}, we now define an upper bound $\bar{\gamma}$ of the submodular ratio $\gamma$ as follows:
\begin{equation}
\bar{\gamma} = \min_{S  \subseteq N}  \frac{ f(\{i\} \mid S)}{ f(\{i\} \mid S \cup \{j\})}, 
\end{equation}
where $i \notin S \cup \{j\}$.
\begin{prop}
\label{prop:aprox_submo}
A function $f(\cdot)$ is approximately submodular if there exists constants $ 1 \ge \bar{\gamma} \geq \gamma > 0$ satisfying any of the following hold:
\begin{description}
  \item[(i)] $f(A) - f(A \cap B) \geq \gamma(f(A \cup B ) -f(B)), \;\; \forall A, B \subseteq N.$
  \item[(ii)] $ f(\{i\} \mid S ) \geq \gamma \; f(\{i\} \mid T ), \;\; \forall S \subseteq T \subseteq N$, $i \notin T.$
  \item[(iii)] $ f(\{i\} \mid S )  \geq \bar{\gamma} \; f(\{i\} \mid S \cup \{ e\}), \;\; \forall S \subseteq N,  i \notin S \cup \{e\}. $
  \item[(iv)]  $f(T) \leq f(S) + f(\{j_1\} \mid S ) + \frac{1}{\bar{\gamma}} \; f(\{j_2\} \mid S ) + \sum_{j \in T \setminus (S \cup \{j_1, j_2\})} {{\frac{1}{ \gamma}}} \; f(\{j\} \mid S ) - \sum_{i \in S \setminus T}\gamma f(\{i\} \mid T \cup S \setminus \{i\} ), \;\; \forall S, T \subseteq N$, $j_1, j_2 \in T \setminus S$.
\item[(v)] $f(T) \leq f(S) + f(\{j_1\} \mid S) + \frac{1}{ \bar{\gamma}} \; f(\{j_2\} \mid S ) + \sum_{j \in T \setminus (S \cup \{j_1, j_2\})} {{\frac{1}{ \gamma}}} \; f(\{j\} \mid S ), \quad \forall S \subseteq T \subseteq N, j_1, j_2 \in T \setminus S$.
\end{description}
\end{prop}

The proof of Proposition~\ref{prop:aprox_submo} is in the Appendix.

\begin{prop}
\label{prop:ndec_aprox_submo}
A function $f(\cdot)$ is non-decreasing approximately submodular if there exists constants $1 \ge \bar{\gamma} \geq \gamma > 0$ satisfying any of the following hold:
\begin{description}
\item[$(\text{i}^{\ast})$]  $f(A) - f(A \cap B) \geq \gamma \; (f(A \cup B ) -f(B)), \;\; \forall A \subseteq B \subseteq N, \; f(A) \leq f(B)$.
\item[$(\text{ii}^{\ast})$] $ f(\{i\} \mid S ) \geq \gamma \; f(\{i\} \mid T ) \geq 0, \;\; \forall S \subseteq T \subseteq N $, $i \notin T.$
\item[$(\text{iv}^{\ast})$] $ f(T) \leq f(S) + f(\{j_1\} \mid S ) + \frac{1}{\bar{\gamma}} \; f(\{j_2\} \mid S ) + \sum_{j \in T \setminus (S \cup \{j_1, j_2\})} \frac{1}{ \gamma} \; f(\{j\} \mid S ), \;\; \forall S,T \subseteq N, j_1, j_2 \in T \setminus S$.
\end{description}
\end{prop}
The proof of Proposition \ref{prop:ndec_aprox_submo} is in the Appendix.
We next consider a set $X$ of $(\eta, \bm{y})$ satisfying the following condition.
\begin{equation}
\begin{array}{l}
\eta \leq f(S) + f(\{j_1\} \mid S ) y_{j_1} + \frac{1}{\bar{\gamma}} \; f(\{j_2\} \mid S) y_{j_2} + \sum_{j \in N \setminus (S \cup \{j_1, j_2\})} \frac{1}{ \gamma} \; f(\{j\}\mid S) y_j, \\
 \forall S \subseteq N, |S| \leq k, j_1, j_2 \in N \setminus S, 
\end{array}
\end{equation}
where $\bm{y} = \{y_{1} ,\dots, y_{n}\} \in \{0,1\}^n$.
\begin{prop}
\label{prop:bin_vec}
Suppose $f(\cdot)$ is a non-decreasing approximately submodular function, $(\eta, \bm{y})$ $\in X$ if and only if $\eta \leq f(T)$, $T = \{ j \in N \mid y_j = 1 \} \subseteq N$.
\end{prop}
The proof of Proposition~\ref{prop:bin_vec} is in the Appendix.
We now replace $\bar{\gamma}$ with $\gamma$ due to $\bar{\gamma} \geq \gamma$. 
We formulate the ASFM problem into the following BIP problem (\ref{eq:apxBIP}).
\begin{equation}
\label{eq:apxBIP}
\begin{array}{ll}
\maximize & z \\
\subjectto & z \leq f(S) + f(\{j\} \mid S) y_{j} + \displaystyle\sum_{i \in N \setminus (S \cup \{j\})} \frac{1}{ \gamma} \; f(\{i\} \mid S)y_i,\\
& j \in N \setminus S,~ S \in F,\\
& \displaystyle\sum_{i \in N} y_{i} \leq k, \\
& y_{i} \in \{0, 1\}, \ i \in N,\\
\end{array}
\end{equation}
where $F$ denotes the set of all feasible solutions satisfying the cardinality constraint $|S| \leq k$.

\section{PROPOSED ALGORITHMS\label{sec:proposed}}
We first present a modified constraint generation algorithm for the ASFM problem based on the algorithm~\citep{Nemhauser1981} in Subsection~\ref{ssec:cg}. 
The modified constraint generation algorithm often needs to solve a large number of reduced BIP problems because of generating only one constraint at each iteration. 
We accordingly propose an improved constraint generation algorithm to generate a promising set of constraints for attaining good upper bounds while solving a smaller number of reduced BIP problems in Subsection~\ref{sec:icg}. Moreover, we develop a branch-and-cut algorithm by using the above algorithm in Subsection~\ref{sec:branch_and_cut}.

\subsection{Modified Constraint Generation Algorithm\label{sec:mcg}}
We first define BIP($Q$) as the following reduced BIP problem of the problem~(\ref{eq:apxBIP}).
\begin{equation}
\label{eq:apr_BIPQ}
\begin{array}{ll}
\maximize & z \\
\subjectto & z \leq f(S) + f(\{j\} \mid S) y_{j} + \displaystyle\sum_{i \in N \setminus (S \cup \{j\})} \frac{1}{ \gamma} \; f(\{i\} \mid S)y_i,\\
& j \in N \setminus S, \; S \in Q,\\
& \displaystyle\sum_{i \in N} y_{i} \leq k,  \\
&  y_{i} \in \{0, 1\}, \; i \in N,
\end{array}
\end{equation}
where $j = \mathrm{argmax}_{i \in N \setminus S } f( \{i\} \mid S)$.
We propose a modified constraint generation algorithm for the ASFM problem based on the constraint generation algorithm for the SFM problem (Subsection \ref{ssec:cg}), where the proposed algorithm solves the above problem (\ref{eq:apr_BIPQ}) instead of (\ref{eq:BIPQ}).

\subsection{Improved Constraint Generation Algorithm\label{sec:icg}}
Let $\boldsymbol{y}^{(t)} = (y_1^{(t)},\dots,y_n^{(t)})$ and $z^{(t)}$ be an optimal solution and the optimal value of $\textnormal{BIP}(Q)$ at the $t$-th iteration of the constraint generation algorithm, respectively.
We note that $z^{(t)}$ gives an upper bound of the optimal value of the problem (\ref{eq:apxBIP}).
To improve the upper bound $z^{(t)}$, it is necessary to add a new feasible solution $S^{\prime} \in F$ to $Q$ satisfying the following inequality.
\begin{equation}
z^{(t)} > f(S^{\prime})  + f(\{j\} \mid S^{\prime}) y_{j}^{(t)}  +  \displaystyle\sum_{i \in N \setminus (S^{\prime} \cup \{j\})} {\frac{1}{ \gamma}} f(\{i\} \mid S^{\prime})\; y_i^{(t)}.
\end{equation}
For this purpose, we now consider the following problem to generate a new feasible solution $S^{\prime} \in F$ adding to $Q$ called the separation problem.
\begin{equation}
\label{eq:pricing}
\begin{array}{ll}
\textnormal{minimize}
& f(S)  + f(\{j\} \mid S) y_{j}^{(t)}  +  \displaystyle\sum_{i \in N \setminus (S \cup \{j\})} {\frac{1}{ \gamma}} f(\{i\} \mid S)\; y_i^{(t)} \\
\textnormal{subject to} & |S| \leq k, \; S \subseteq N, j \in N \setminus S. 
\end{array}
\end{equation}
If the optimal value of the separation problem (\ref{eq:pricing}) is less than $z^{(t)}$, then we add an optimal solution $S^{\prime}$ of the separation problem (\ref{eq:pricing}) to $Q$; otherwise, we conclude $z^{(t)}$ is the optimal value of the problem (\ref{eq:apxBIP}).
We repeat adding a new feasible solution $S^{\prime}$ obtained from the separation problem (\ref{eq:pricing}) to $Q$ and solving the updated $\textnormal{BIP}(Q)$ until $z^{(t)}$ and $f(S^{\prime})$ meet.
This procedure is often called the cutting-plane algorithm which is used for the mixed integer programs~\citep{Marchand2002}.
However, the computational cost to solve a separation problem (\ref{eq:pricing}) is very expensive, almost the same as solving the SFM problem.
To overcome this, we propose an improved constraint generation algorithm to quickly generate a promising set of constraints. 

After solving $\textnormal{BIP}(Q)$, we obtain at least one feasible solution $S^{\natural} \in Q$ attaining the optimal value $z^{(t)}$ of $\textnormal{BIP}(Q)$, i.e.,
\begin{equation}
\label{eq:tight}
z^{(t)} = f(S^{\natural}) + f(\{j\}) \mid S^{\natural})\; y_j^{(t)} + \displaystyle\sum_{i \in N \setminus (S^{\natural}\cup \{j\}) } \frac{1}{\gamma}~f(\{i\} \mid S^{\natural} ) \; y_i^{(t)}.
\end{equation}
Let $S^{(t)}$ be the optimal solution of $\textnormal{BIP}(Q)$ corresponding to $\boldsymbol{y}^{(t)}$, where we assume $S^{(t)} \not\in Q$.

We then consider adding an element $j \in S^{(t)} \setminus S^{\natural}$ to $S^{\natural}$. 
In the case with satisfying $j = \mathrm{argmax}_{i \in N \setminus S } f( \{i\} \mid S)$, we obtain the following inequality by approximately submodularity:
\small
\begin{equation}
\label{eq:add}
\begin{array}{lll}
z^{(t)} &=& f(S^{\natural}) + f(\{j\}) \mid S^{\natural})  \; y_j^{(t)}+ \displaystyle\sum_{i \in N \setminus (S^{\natural} \cup \{j\}) } \frac{1}{\gamma}~ f(\{i\} \mid S^{\natural}) \; y_i^{(t)}\\
&=& f(  S^{\natural} \cup \{ j \} )  + \displaystyle\sum_{i \in N \setminus (S^{\natural} \cup \{ j \}) } \frac{1}{\gamma}~ f(\{i\} \mid S^{\natural}) \; y_i^{(t)}\\
& \geq &  f(S^{\natural} \cup \{j\}) + \displaystyle\sum_{i \in N \setminus (S^{\natural} \cup \{j\})} f( \{i\} \mid S^{\natural} \cup \{j\}) \; y_i^{(t)},
\end{array}
\end{equation}
\normalsize
where $y_j^{(t)} = 1$ due to $j \in S^{(t)}$.
In the other case when $j \neq \mathrm{argmax}_{i \in N \setminus S } f( \{i\} \mid S) $, we obtain a similar inequality above.
Ideally, we should have $1 / \gamma$ in the sum of right hand side of the inequality.
By the inequality (\ref{eq:add}) with $\gamma \simeq 1$, we observe that it is preferable to add the element $j \in S^{(t)} \setminus S^{\natural}$ to $S^{\natural}$ for improving the upper bound $z^{(t)}$.
Here, we note that it is necessary to remove another element $i \in S^{\natural}$ if $|S^{\natural}| = k$ holds.

Based on this observation, we develop a heuristic algorithm to generate a set of new feasible solutions $S^{\prime} \in F$ for improving the upper bound $z^{(t)}$.
Given a set of feasible solutions $Q \subseteq F$, let $q_i$ be the number of feasible solutions $S \in Q$ including an element $i \in N$.
We define the occurrence rate $p_i$ of each element $i$ with respect to $Q$ as 
\begin{equation}
p_i = \frac{ q_i }{ \sum_{ j \in N}{ q_j}}.
\end{equation}
For each element $i \in S^{\natural} \cup S^{(t)}$, we set a random value $r_i$ satisfying $0 \leq r_i \leq p_i$.
If there are multiple feasible solutions $S^{\natural} \in Q$ satisfying the equation (\ref{eq:tight}), then we select one of them at random.
We take the $k$ largest elements $i \in S^{\natural} \cup S^{(t)}$ with respect to the value $r_i$ to generate a feasible solution $S^{\prime} \in F$.
\begin{description}
\item[\underline{\textbf{Algorithm SUB-ICG}$(Q, S^{(t)}, \lambda)$}]
\item[\textbf{Input:}]~ A set of feasible solutions $Q \subseteq F$.
A feasible solution $S^{(t)} \not\in Q$.
The number of feasible solutions to be generated $\lambda$.
\item[\textbf{Output:}]~ A set of feasible solutions $Q^{\prime} \subseteq F$.
\item[\textbf{Step1:}]~ Set $Q^{\prime} \leftarrow \emptyset$ and $h \leftarrow 1$.
\item[\textbf{Step2:}]~ Select a feasible solution $S^{\natural} \in Q$ satisfying the equation (\ref{eq:tight}) at random. Set a random value $r_i$ $(0 \le r_i \le p_i)$ for $i \in S^{\natural} \cup S^{(t)}$. 
\item[\textbf{Step3:}]~ If $|S^{\natural}| = k$ holds, then take the $k$ largest elements $i \in S^{\natural} \cup S^{(t)}$ with respect to $r_i$ to generate a feasible solution $S^{\prime} \in F$.
Otherwise, take the largest element $i \in S^{(t)} \setminus S^{\natural} $ with respect to $r_i$ to generate a feasible solution $S^{\prime} = S^{\natural} \cup \{i\} \in F$.
\item[\textbf{Step4:}]~ If $S^{\prime} \not\in Q^{\prime}$ holds, then set $Q^{\prime} \leftarrow Q^{\prime} \cup \{ S^{\prime} \}$ and $h \leftarrow h + 1$.
\item[\textbf{Step5:}]~ If $h = \lambda$ holds, then output $Q^{\prime}$ and exit.
Otherwise, return to Step2.
\end{description}

We summarize the improved constraint generation algorithm as follows, in which we define $Q$ as the set of feasible solutions $S^{(0)}, S^{(1)}, \dots, S^{(t-1)}$ obtained by solving reduced BIP problems and $Q^+$ as the set of feasible solutions generated by $\textnormal{SUB-ICG}(Q, S^{(t)}, \lambda)$.

\begin{description}
\item[\underline{\textbf{Algorithm ICG}$(S^{(0)},\lambda)$}]
\item[\textbf{Input:}]~ The initial feasible solution $S^{(0)}$.
The number of feasible solutions to be generated at each iteration $\lambda$.
\item[\textbf{Output:}]~ The incumbent solution $S^{\ast}$.
\item[\textbf{Step1:}]~ Set $Q \leftarrow \{ S^{(0)}$\}, $Q^{+} \leftarrow \{ S_{[0]}^{(0)},\dots, S_{[k]}^{(0)} \}$, $S^{\ast} \leftarrow S^{(0)}$ and $t \leftarrow 1$.
\item[\textbf{Step2:}]~ Solve $\textnormal{BIP}(Q^{+})$.
Let $S^{(t)}$ and $z^{(t)}$ be an optimal solution and the optimal value of $\textnormal{BIP}(Q^{+})$, respectively.
\item[\textbf{Step3:}]~ If $f(S^{(t)}) > f(S^{\ast})$ holds, then set $S^{\ast} \leftarrow S^{(t)}$.
\item[\textbf{Step4:}]~ If $z^{(t)} = f(S^{\ast})$ holds, then output the incumbent solution $S^{\ast}$ and exit.
\item[\textbf{Step5:}]~ Set $Q \leftarrow Q \cup \{ S^{(t)} \}$, $Q^{+} \leftarrow Q^{+}$ $\cup$  $\{ S^{(t)} \}$ $\cup$ $\textnormal{SUB-ICG}(Q,S^{(t)}, \lambda)$ and  $t \leftarrow t + 1$. 
\item[\textbf{Step6:}]~ For each feasible solution $S \in \textnormal{SUB-ICG}(Q,S^{(t)},\lambda)$, if $f(S) > f(S^{\ast})$ holds, then set $S^{\ast} \leftarrow S$. 
Return to Step2.
\end{description}

We note that the improved constraint generation algorithm often attains good lower bounds as well as the upper bounds because SUB-ICG gives good feasible solutions at each iteration.
\subsection{Branch-and-Cut Algorithm\label{sec:branch_and_cut}}
We propose a branch-and-cut algorithm incorporating the improved constraint generation algorithm.
We first define the search tree of the branch-and-cut algorithm. 
Each node $(S^0, S^1)$ of the search tree consists of a pair of sets $S^0$ and $S^1$, where elements $i \in S^0$ (resp., $i \in S^1$) correspond to variables fixed to $y_i = 0$ (resp., $y_i = 1$) of the problem (\ref{eq:apxBIP}). 
The root node is set to $(S^0, S^1) \leftarrow (\emptyset, \emptyset)$.
Each node $(S^0,S^1)$ has two children $(S^0 \cup \{ i^{\ast} \},S^1)$ and $(S^0, S^1 \cup \{ i^{\ast} \})$, where $i^{\ast} = \mathrm{argmax}_{i \in N \setminus (S^0 \cup S^1)} f(S^1 \cup \{ i \})$.

The branch-and-cut algorithm employs a stack list $L$ to manage nodes of the search tree.
The value of a node $(S^0,S^1)$ is defined as the optimal value $z^{(S^0,S^1)}$ of the following reduced BIP problem $\textnormal{BIP}(Q^+, S^0, S^1)$:
\begin{equation}
\begin{array}{lll}
\textnormal{maximize} & z\\
\textnormal{subject to} & \multicolumn{2}{l}{z \le f(S) + f(\{j\} \mid S) y_{j} + \displaystyle\sum_{i \in N \setminus (S \cup \{j\})} {\frac{1}{ \gamma}} f(\{i\} \mid S)y_i,}\\
& \multicolumn{2}{l}{S \in Q^+, \; j \in N \setminus S,}\\
& \multicolumn{2}{l}{\displaystyle\sum_{i \in N \setminus (S^0 \cup S^1)} y_i \le k - |S^1|,}\\
& y_i \in \{ 0, 1 \}, & i \in N \setminus (S^0 \cup S^1),\\
& y_i = 0, & i \in S^0,\\
& y_i = 1, & i \in S^1,
\end{array}
\end{equation}
where $j = \mathrm{argmax}_{i \in N \setminus S } f( \{i\} \mid S)$, and $Q^+$ is the set of feasible solution generated by the improved constraint generation algorithm so far.
We note that $z^{(S^0,S^1)}$ gives an upper bound of the optimal value of the problem (\ref{eq:apxBIP}) at the node $(S^0,S^1)$; i.e., under the condition that $y_i = 0$ $(i \in S^0)$ and $y_i = 1$ $(i \in S^1)$.

We start with a pair of sets $Q = \{ S \}$ and $Q^+ = \{ S_{[0]},\dots, S_{[k]} \}$, where $S$ is the initial feasible solutions obtained by the greedy algorithm \citep{Minoux1978,Nemhauser1978}.
To obtain good upper and lower bounds quickly, we first apply the first $k$ iterations of the improved constraint generation algorithm.
We then repeat to extract a node $(S^0,S^1)$ from the top of the stack list $L$ and insert its children into the top of the stack list $L$ at each iteration.
Thus, we employ a depth-first-search for the tree search of the branch-and-cut algorithm.

Let $(S^0,S^1)$ be a node extracted from the stack list $L$, and $S^{\ast}$ be the incumbent solution of the problem (\ref{eq:apxBIP}) (i.e., the best feasible solution  obtained so far).
We first solve $\textnormal{BIP}(Q^+, S^0, S^1)$ to obtain an optimal solution $S^{(S^0,S^1)}$ and the optimal value $z^{(S^0,S^1)}$.
We then generate a set of feasible solutions by $\textnormal{SUB-ICG}(Q,S^{(S^0,S^1)},\lambda)$.
For each feasible solution $S^{\prime} \in \{ S^{(S^0,S^1)} \} \cup \textnormal{SUB-ICG}(Q,S^{(S^0,S^1)},\lambda)$, if $f(S^{\prime}) > f(S^{\ast})$ holds, then we replace the incumbent solution $S^{\ast}$ with $S^{\prime}$.
If $z^{(S^0,S^1)} > f(S^{\ast})$ holds, then we insert the two children $(S^0 \cup \{ i^{\ast} \}, S^1)$ and $(S^0, S^1 \cup \{ i^{\ast} \})$ into the top of the stack list $L$ in this order.

To decrease the number of reduced BIP problems to be solved in the branch-and-cut algorithm, we keep the optimal value $z^{(S^0,S^1)}$ of $\textnormal{BIP}(Q^+,S^0,S^1)$ as an upper bound $\bar{z}^{(S^0 \cup \{ i^{\ast} \}, S^1)}$ (resp., $\bar{z}^{(S^0,S^1 \cup \{ i^{\ast} \})}$) of the child $(S^0 \cup \{ i^{\ast} \}, S^1)$ (resp., $(S^0, S^1 \cup \{ i^{\ast} \})$) when inserted to the stack list $L$.
If $\bar{z}^{(S^0,S^1)} \le f(S^{\ast})$ holds when we extract a node $(S^0,S^1)$ from the stack list $L$, then we can prune the node $(S^0,S^1)$ without solving $\textnormal{BIP}(Q^+,S^0,S^1)$.
We set the upper bound $\bar{z}^{(\emptyset,\emptyset)}$ of the root node $(\emptyset,\emptyset)$ to $\infty$. 
We repeat these procedures until the stack list $L$ becomes empty.
\smallskip
\begin{description}
\item[\underline{\textbf{Algorithm BC-ICG}$(S,\lambda)$}]
\item[\textbf{Input:}]~ The initial feasible solution $S$.
The number of feasible solutions to be generated at each node $\lambda$. 
\item[\textbf{Output:}]~ The incumbent solution $S^{\ast}$.
\item[\textbf{Step1:}]~ Set $L \leftarrow \{ (\emptyset,\emptyset) \}$, $\bar{z}^{(\emptyset,\emptyset)} \leftarrow \infty$, $Q \leftarrow \{ S\}$,  $Q^+ \leftarrow\{ S_{[0]},\dots, S_{[k]} \}$ and $S^{\ast} \leftarrow S$.
\item[\textbf{Step2:}]~ Apply the first $k$ iterations of $\textnormal{ICG}(S,\lambda)$ to update the sets $Q$ and $Q^+$ and the incumbent solution $S^{\ast}$.
\item[\textbf{Step3:}]~ If $L = \emptyset$ holds, then output the incumbent solution $S^{\ast}$ and exit.
\item[\textbf{Step4:}]~ Extract a node $(S^0,S^1)$ from the top of the stack list $L$.
If $\bar{z}^{(S^0,S^1)} \le f(S^{\ast})$ holds, then return to Step3.
\item[\textbf{Step5:}]~ Solve $\textnormal{BIP}(Q^+,S^0,S^1)$.
Let $S^{(S^0,S^1)}$ and $z^{(S^0,S^1)}$ be an optimal solution and the optimal value of $\textnormal{BIP}(Q^+,S^0,S^1)$, respectively.
\item[\textbf{Step6:}]~ Set $Q \leftarrow Q \cup \{ S^{(S^0,S^1)} \}$, $Q^+ \leftarrow Q^+ \cup \{ S^{(S^0,S^1)} \} \cup \textnormal{SUB-ICG}(Q,S^{(S^0,S^1)},\lambda)$.
\item[\textbf{Step7:}]~ For each feasible solution $S^{\prime} \in \{ S^{(S^0,S^1)} \} \cup \textnormal{SUB-ICG}(Q,S^{(S^0,S^1)},\lambda)$, if $f(S^{\prime}) > f(S^{\ast})$ holds, then set $ S^{\ast} \leftarrow  S$.
\item[\textbf{Step8:}]~ If $ z^{(S^0,S^1)} \leq f(S^{\ast})$, then return to Step3.
\item[\textbf{Step9:}]~ If $|S^0 \cup S^1| \le n-1$ and $|S^1| \le k-1$ hold, then set $L \leftarrow L \cup \{ (S^0 \cup \{ i^{\ast} \}, S^1), (S^0, S^1 \cup \{ i^{\ast} \}) \}$, $\bar{z}^{(S^0 \cup \{ i^{\ast} \}, S^1)} \leftarrow z^{(S^0,S^1)}$ and $\bar{z}^{(S^0, S^1\cup \{ i^{\ast} \})} \leftarrow z^{(S^0,S^1)}$, where $i^{\ast} = \mathrm{argmax}_{i \in N \setminus (S^0 \cup S^1)} \allowbreak 
f(S^1 \cup \{ i \})$.
Return to Step3.
\end{description}
\smallskip

We note that the branch-and-cut algorithm is similar to that for the traveling salesman problem based on a BIP formulation with an exponential number of subtour elimination constraints \citep{Crowder1980,Grotschel1991}.

\section{EXAMPLE}
\label{sec:example}
We consider the combinatorial auction (CA) problem that asks a bidder to select a package of items to maximize its utility, which is formulated as the ASFM problem.
A greedy algorithm obtains a feasible solution quickly, however it often fails to obtain important items for the problem.
To do so, we need an optimal solution as well as other good solutions whose objective value are close to the optimal value.
For this purpose, we modify BC-ICG to hold all incumbent solutions obtained so far as well as the final incumbent solution.

\noindent\textbf{ Combinatorial auction (CA)} 
We are given a set of $n$ items $N = \{1,\dots,n\}$.
We select a set of items $S \subseteq N$ to make a package of items.
We define $w_i$ as the individual utility of an item $i \in N$ and $r_{ij}$ as the mutual utility for a pair of items $i,j \in N$.
The utility of a package of items is defined as 
\begin{equation}
f(S) = \displaystyle\sum_{ i \in S} w_i + \displaystyle\sum_{ i,j \in S} r_{ij}. 
\end{equation}
We have tested BC-ICG for an instance arising from a supermarket transaction data containing 170 items and 9835 transactions.\footnote{\url{http://www.sci.csueastbay.edu/~esuess/classes/Statistics_6620/Presentations/ml13/groceries.csv}}
We set the size of the package $k = 2,3$.
We also set the individual utility randomly $w_i \in [1,2]$, while setting the mutual utility $r_{ij} \in [-0.09, 0.01]$ according to the number of times that both items $i,j$ are selected in the same transaction.
We obtain a lower bound $\underline{\gamma}$ of the submodular ratio $\gamma$ by the following formula:
\begin{equation}
\underline{\gamma} = \min_{i \in N} \Biggl\{ \frac{   w_i + \sum_{j \in L_{[q_1]}} r_{ij}  } { w_i + \sum_{j \in U_{[q_2]}} r_{ij}  } \Biggr\}.
\end{equation}
where the sets $U$ and $L$ are the non-increasing positive and non-decreasing negative ordered set with respect to $r_{ij}$ for an item $i \in N$, respectively, and $q_1 = \min \{k-1, |L|\}$, $q_2 = \min \{k-1, |U|\}$.
We note that $L_{[q_1]}$ (resp., $U_{[q_2]}$) represents the first $q_1$ (resp., $q_2$) elements of a sorted set $L$ (resp., $U$).
For the instance, we obtained $\underline{\gamma} = 0.906, 0.736$ with $k=2,3$, respectively.  

Table \ref{tab:CA} shows the frequency of items in the series of solutions obtained by BC-ICG.
The optimal solutions obtained by BC-ICG are [``yogurt'', ``frozen vegetables''], [``yogurt'', ``sugar'', ``organic products''] for $k=2,3$, respectively.
On the other hand, the feasible solutions obtained the greedy algorithm are [``tea'', ``yogurt''], [``tea'', ``yogurt'', ``frozen vegetables''] for $k = 2, 3$, respectively.
We note that ``tea'' is not selected in the optimal solutions while it has the largest value of the 170 items.
That is, the greedy algorithm sometimes fails to attain important items constitute the optimal solution.

\begin{table}[t]
\caption{Frequency of items in a series of solutions obtained by BC-ICG.}
\centering
\scalebox{0.70}{
{\small
\begin{tabular}{@{}r@{~}r@{~}r@{~}r@{~}r@{~}r@{\,~~}r@{\,~~}r@{~}r@{}}
 \hline
 \multicolumn{1}{c}{} &
\multicolumn{1}{c}{``tea''} & \multicolumn{1}{c}{``yogurt''} & \multicolumn{1}{c}{``sugar''} & \multicolumn{1}{c}{``organic products''} & \multicolumn{1}{c}{``frozen vegetables''} &  \multicolumn{1}{c}{``softner''}       \\
       \hline
$k=2$ & 1 & 3  &  2 & 1   & 2 & 1   \\
  \hline
$k=3$ & 1 & 4 & 2 & 2 & 3 & 0 \\
\hline
 \end{tabular}
}
}
 \label{tab:CA}
\end{table}


\section{COMPUTATIONAL RESULTS}
\label{sec:results}
We tested two existing algorithms:~(i)~the $\textnormal{A}^{\ast}$~search algorithm with the heuristic function $h_{mod}$ ($\textnormal{A}^{\ast}$-MOD),  (ii)~the modified constraint generation algorithm (MCG) and two proposed algorithms:~(iii)~the improved constraint generation algorithm (ICG), (iv)~the branch-and-cut algorithm (BC-ICG).
All algorithms were tested on a personal computer with a 4.0~GHz Intel Core i7 processor and 32~GB memory.
For MCG, ICG, and BC-ICG, we use an mixed integer programming (MIP) solver called CPLEX~12.8~(\citeyear{CPLEX2019}) for solving reduced BIP problems, and the number of feasible solutions to be generated at each iteration $\lambda$ is set to $10k$ based on computational results of preliminary experiments.

We report computational results for three types of well-known benchmark instances called \textit{facility location} (LOC), \textit{weighted coverage} (COV), and \textit{bipartite influence} (INF) according to~\cite{Kawahara2009} and~\cite{Sakaue2018}.
We note that those instances were originally generated for the SFM problem.
For generating instances for the ASFM problem, we replace the original utility function $f(S)$ with $f(S) + r_S$ for a number of feasible solutions, where $r_S$ is a reward value for a feasible solution $S \subseteq F$.
We randomly selected $1000k$ feasible solutions $S \subseteq F$ and modified their utility functions while satisfying non-decreasing and a given lower bound $\underline{\gamma} = 0.8$ of the submodular ratio $\gamma$.

\smallskip
\noindent\textbf{Facility location (LOC)} 
We are given a set of $n$ locations $N = \{ 1,\dots,n \}$ and a set of $m$ clients $M = \{ 1,\dots,m \}$.
We consider to select a set of $k$ locations to build facilities.
We define $g_{ij} \geq 0$ as the benefit of a client $i \in M$ attaining from a facility of location $j \in N$.
We select a set of locations $S \subseteq N$ to built the facilities.
Each client $i \in M$ attains the benefit from the most beneficial facility.
The total benefit for the clients is defined as 
\begin{equation}
f(S) = \displaystyle\sum_{i \in M} \max_{j \in S} g_{ij}. 
\end{equation}

\smallskip
\noindent\textbf{Weighted coverage (COV)}
We are given a set of $m$ items $M = \{ 1,\dots,m \}$ and a set of $n$ sensors $N = \{ 1,\dots,n \}$.
Let $M_j \subseteq M$ be the subset of items covered by a sensor $j \in N$, and $w_i \ge 0$ be a weight of an item $i \in M$.
We select a set of sensors $S \subseteq N$ to cover items.
The total weighted coverage for the items is defined as
\begin{equation}
f(S) = \sum_{i \in M} w_i \max_{j \in S} a_{ij},
\end{equation}
where $a_{ij} = 1$ if $i \in M_j$ holds and $a_{ij} = 0$ otherwise.

\smallskip
\noindent\textbf{Bipartite influence (INF)}
We are given a set of $m$ targets $M = \{ 1, \dots, m \}$ and a set of items $N = \{ 1, \dots, n \}$.
Given a bipartite graph $G = (M, N; A)$, where $A \subseteq M \times N$ is a set of directed edges, we consider an influence maximization problem on $G$.
Let $p_j \in [0,1]$ be the activation probability of an item $j \in N$.
The probability that a target $i \in M$ gets activated by a set of items $S \subseteq N$ is $1 -  \prod_{j \in S} (1 - q_{ij})$, where  $q_{ij} = p_j$ if $(i,j) \in A$ holds and $q_{ij} = 0$ otherwise. 
We select a set of items $S \subseteq N$ to activate targets.
The expected number of targets activated by a set of items $S \subseteq N$ is defined as
\begin{equation}
f(S) = \displaystyle\sum_{i \in M} \left( 1- \displaystyle\prod_{j \in S} (1 - q_{ij}) \right).
\end{equation}
\smallskip

We tested all algorithms for 18 classes of randomly generated instances that are characterized by several parameters.
We set $m = n + 1$ and $k = 5,8$ for LOC, COV and INF instances according to~\citep{Kawahara2009}.
We set $n = 20,30,40$ for LOC instances and $n = 20,40,60$ for COV and INF instances.
For LOC instances, $g_{ij}$ is a random value taken from interval $[0,1]$.
For COV instances, a sensor $j \in N$ randomly covers an item $i \in M$ with probability $0.15$, and $w_i$ is a random value taken from interval $[0,1]$.
For INF instances, $p_j$ is a random value taken from interval $[0,1]$, and the bipartite graph $G$ is a random graph in which an edge $(i,j) \in A$ is generated randomly with probability $0.1$.
We set these parameters to different values from those in~\citep{Sakaue2018} considering the difference between the cardinality and knapsack constraints. 
For each class of instances, five instances were generated and tested.
For all instances, we set the time limit to 7200 seconds.

\begin{figure}[tb]
  \centering
  \includegraphics[height=0.35\textheight]{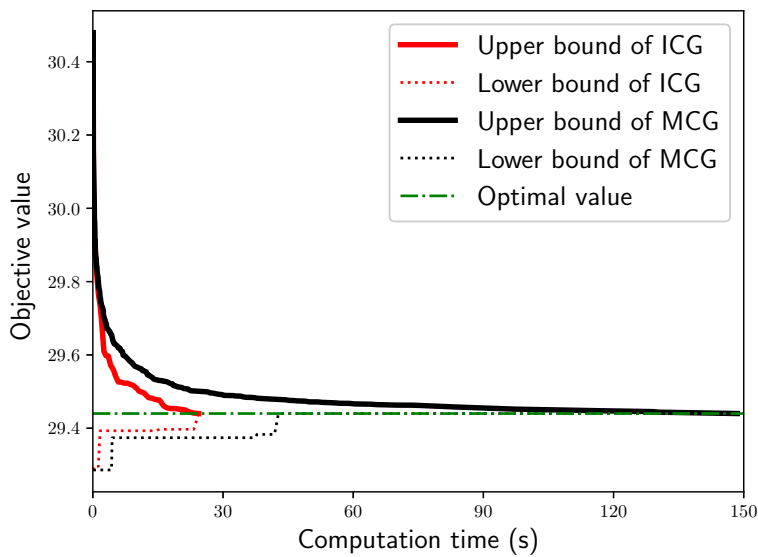}
  \caption{Trends of the upper and lower bounds obtained by MCG and ICG with respect to the elapsed computation time.}
  \label{fig:UBLB}
\end{figure}

Tables~\ref{tab:time} and \ref{tab:node} show the average computation time (in seconds) and the average number of processed nodes of the algorithms for each class of instances, respectively.
If an algorithm could not solve an instance optimally within the time limit, then we set the computation time to 7200 seconds.
The best computation time among the compared algorithms is highlighted in bold.
The numbers in parentheses show the number of instances optimally solved within the time limit.
According to Table~\ref{tab:time}, ICG performed better than MCG in most of instances.
Figure~\ref{fig:UBLB} shows trends of the upper and lower bounds obtained of MCG and ICG with respect to the computation time for an LOC instance ($n= 30$, $k=8$).
We can see that ICG attained better upper and lower bounds than those of MCG by generating good feasible solutions and adding them as constraints at each iteration.
The number of iterations that MCG and ICG solved a reduced BIP problem were 275 and 35, respectively.
We succeeded in improving the efficiency of MCG by reducing the number of the iterations.
According to Table~\ref{tab:node}, BC-ICG processed much smaller number of nodes than $\textnormal{A}^{\ast}$-MOD due to ICG attaining good upper bounds.
We note that $\textnormal{A}^{\ast}$-MOD performed well for INF instances, because the utility function $f(S)$ became close to linear and $\textnormal{A}^{\ast}$-MOD gave tight upper bound $\bar{f}(S) = f(S) + h(S)$ for the instances.

\begin{figure}[tb]
  \centering
  \includegraphics[height=0.35\textheight]{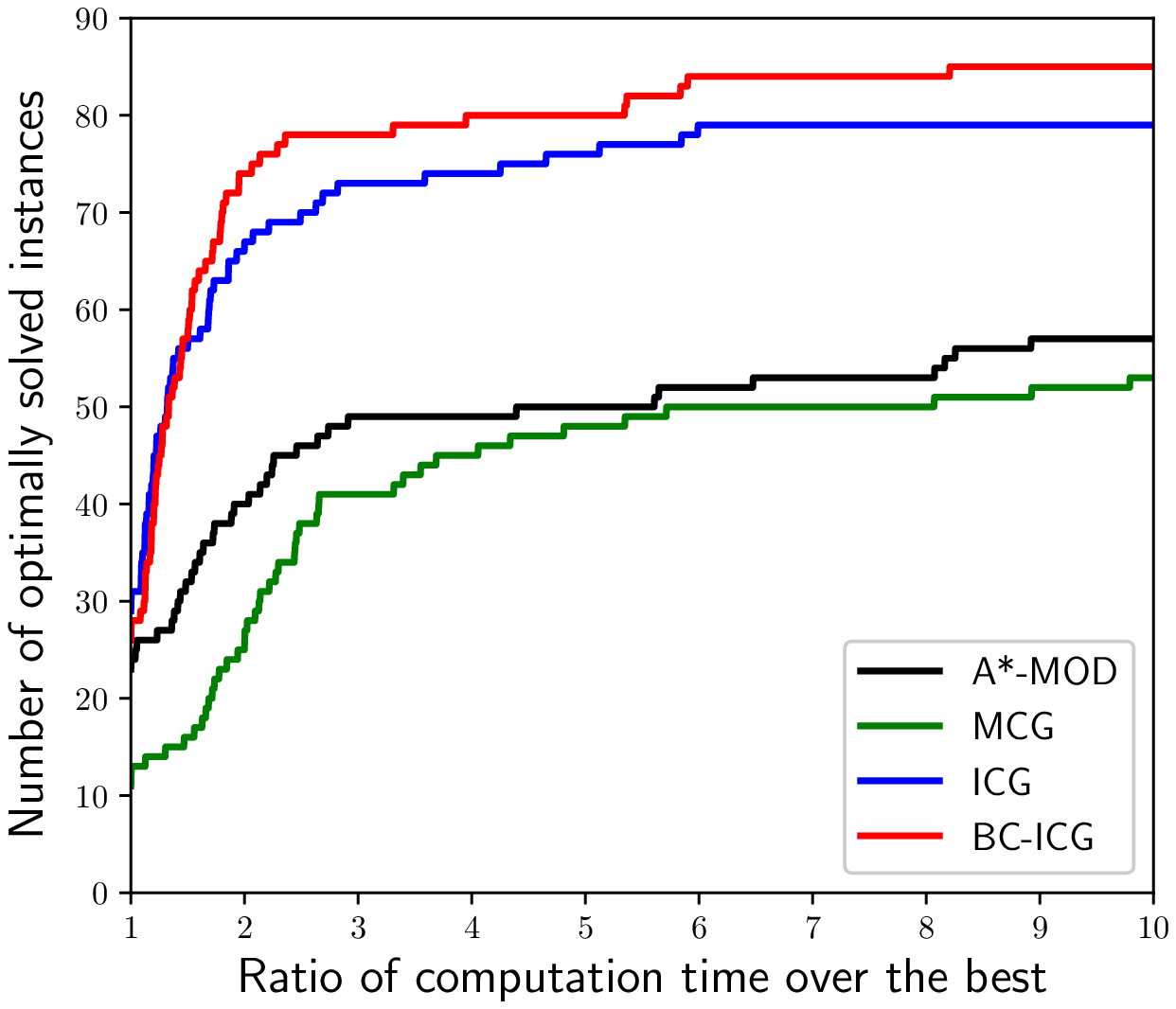}
  \caption{Performance profile for the algorithms.}
  \label{fig:pp}
\end{figure}

Figure~\ref{fig:pp} shows performance profiles~\citep{Dolan2002} of the algorithms for a parameter $1 \le \beta \le 10$.
For given sets of algorithms $\mathcal{A}$ and instances $\mathcal{I}$, the performance profile is defined in terms of computation time $T(A,I)$ of an algorithm $A \in \mathcal{A}$ to solve an instance $I \in \mathcal{I}$ optimally.
For a pair of algorithm $A \in \mathcal{A}$ and instance $I \in \mathcal{I}$, the performance ratio  $R(A,I)$ (i.e., the ratio of computation time over the best) is defined as
\begin{equation}
R(A,I) = \frac{T(A,I)}{\displaystyle\min_{A^{\prime} \in \mathcal{A}} T(A^{\prime},I)},
\end{equation}
where we set $R(A,I) = \infty$ if none of the algorithms solved the instance $I$ optimally.
We note that $R(A,I) \ge 1$ holds by definition.
The performance profile of an algorithm $A \in \mathcal{A}$ illustrates the function $\rho_{\scalebox{0.5}{A}}(\beta)$ that represents the number of instances $I \in \mathcal{I}$ satisfying $R(A,I) \le \beta$.
We observed that BC-ICG solved 78 instances optimally out of all 90 instances while MCG solved only 41 instances with $\beta = 3$.
These computational results show that BC-ICG improved the efficiency of the conventional MCG.
\begin{table}[t]
\caption{Computation time (in seconds) of the proposed and the existing algorithms.}
\centering
\scalebox{0.85}{
{\small
\begin{tabular}{@{}c@{~}c@{~}c@{~}r@{\,~~}r@{\,~~}r@{\,~~}r@{\,~~}r@{\,~~}r@{}}
 \hline
\multicolumn{1}{c}{Type} & \multicolumn{1}{c}{$n$} & \multicolumn{1}{c}{$k$} & \multicolumn{1}{c}{$\underline{\gamma}$} &  \multicolumn{1}{c}{A$^{\ast}$-MOD} & \multicolumn{1}{c}{MCG} & \multicolumn{1}{c}{ICG} & \multicolumn{1}{c}{BC-ICG}       \\
 \hline
      & $20$ & $5$  &  $0.8$ &  0.97 (5) & 1.39 (5)  & $\boldsymbol{ 0.52}$ (5) & 0.70 (5)   \\
   LOC     &  $30$ & $5$  & $0.8$ &  $\boldsymbol{8.69}$ (5) &  27.98 (5) & 11.57 (5) &  10.98 (5)   \\
       &  $40$ & $5$  & $0.8$ &   $\boldsymbol{ 51.00}$ (5)&  2369.62 (5)  & 973.04 (5)  &   183.09 (5)  \\

    \hline
      & $20$ & $8$  &   $0.8$ &  13.22 (5) & 30.01 (5)   &  0.35 (5) &  $\boldsymbol{  0.32}$ (5)   \\
   LOC     &  $30$ & $8$  &   $0.8$  & 409.50 (5)& 99.74 (5)  & 21.77 (5)  &  $\boldsymbol{  17.29}$ (5)    \\
       &  $40$ & $8$  &   $0.8$  & $>$ 5703.85 (4)  & $>$ 5367.68 (3) & $>$ 3433.23 (3) & $\boldsymbol{ 667.62}$ (5)  \\
       \hline

      & $20$ & $5$  &   $0.8$ &  0.40 (5)&  0.36 (5)  & $\boldsymbol{ 0.25}$ (5) &  0.29 (5)   \\
   COV    &  $40$ & $5$  & $0.8$ & 16.36 (5) & 52.49 (5) & $\boldsymbol{ 9.98}$ (5) &  14.15 (5) \\
       &  $60$ & $5$  &   $0.8$  &  163.36 (5)& 1739.64 (5)  & 262.39 (5)   & $\boldsymbol{ 145.73}$ (5)  \\
       \hline
         & $20$ & $8$  &  $0.8$ &  8.71 (5) & $\boldsymbol{ 0.07}$ (5)  & 0.10 (5)  & 0.09 (5) \\
COV  & $40$ & $8$  &  $0.8$ & $>$ 3468.47 (4)  & $>$ 1441.96 (4) & $\boldsymbol{ 2.31}$ (5)& 3.18 (5)\\
    &  $60$ & $8$ &  $0.8$ & $>$ 7200.00 (0) & 108.86 (5)   &  35.81 (5) &  $\boldsymbol{ 32.49}$ (5)  \\
 \hline
   & $20$ & $5$  &  $0.8$  &  $\boldsymbol{ 0.36}$ (5) &  2.46 (5)  & 0.61 (5)  &  1.07 (5)  \\
INF  & $40$ & $5$  &  $0.8$  & $\boldsymbol{ 8.61}$ (5)  & $>$ 1992.35 (4)  &  15.98 (5) &   22.14 (5) \\
    &  $60$ & $5$ &  $0.8$  &   61.20 (5) & $>$ 4467.06 (2) &  $\boldsymbol{ 16.68 }$ (5) &   25.17 (5)  \\
    \hline
   & $20$ & $8$  &  $0.8$  &  3.32 (5) &  3.46 (5)  & 15.98 (5) & $\boldsymbol{ 1.75}$ (5)    \\
INF  & $40$ & $8$  &  $0.8$  & $\boldsymbol{ 438.83}$ (5) & $>$ 3775.80 (3) & $>$ 1670.35 (4)  &  912.62 (5)  \\
    &  $60$ & $8$ &  $0.8$  & $>$ $\boldsymbol{ 4980.66}$ (4) & $>$ 7200.00 (0)  & $>$ 7200.00 (0)  &  $>$ 7200.00 (0)   \\

\hline
 \end{tabular}
}
}
 \label{tab:time}
\end{table}
\begin{table}[t]
\caption{Number of processed nodes by the proposed and the existing algorithms.}
\centering
\scalebox{0.85}{
{\small
\begin{tabular}{@{}c@{~}c@{~}c@{~}r@{\,~~}r@{\,~~}r@{\,~~}r@{}}
 \hline
\multicolumn{1}{c}{Type} & \multicolumn{1}{c}{$n$} & \multicolumn{1}{c}{$k$} & \multicolumn{1}{c}{$\underline{\gamma}$} & \multicolumn{1}{c}{A$^{\ast}$-MOD} &  \multicolumn{1}{c}{BC-ICG}       \\

       \hline
      & $20$ & $5$  &   $0.8$ &  $5.62 \times 10^3$ (5)  & $8.60 \times 10^0$ (5)  \\
   LOC     &  $30$ & $5$  &   $0.8$  & $2.74 \times 10^4$ (5) &    $8.30 \times 10^1$  (5)  \\
       &  $40$ & $5$  &   $0.8$ &  $1.03 \times 10^5$ (5) &  $3.77 \times 10^2$ (5)\\

    \hline
      & $20$ & $8$  &   $0.8$ &  $4.68 \times 10^4$ (5) &  $1.00 \times 10^0$  (5) \\
   LOC     &  $30$ & $8$  &   $0.8$  & $6.55 \times 10^5$ (5) &    $4.78 \times 10^1$  (5) \\
       &  $40$ & $8$  &   $0.8$  & $>$ $3.71 \times 10^6$ (4) &  $3.09 \times 10^2$  (5) \\
       \hline
      & $20$ & $5$  &   $0.8$ &  $2.03 \times 10^3$ (5)  &  $1.00 \times 10^0$  (5) \\
   COV    &  $40$ & $5$  &   $0.8$  & $2.63 \times 10^4$ (5)   &  $6.14 \times 10^1$  (5)\\
       &  $60$ & $5$  &   $0.8$  &  $1.07 \times 10^5$ (5)  & $1.96 \times 10^2$ (5) \\
       \hline
         & $20$ & $8$  &  $0.8$ & $3.54 \times 10^4$ (5)   &  $1.00 \times 10^0$  (5)\\
COV  & $40$ & $8$  &  $0.8$ & $>$ $2.71 \times 10^6$ (4)  & $1.00 \times 10^0$ (5) \\
    &  $60$ & $8$ &  $0.8$ &  $>$ $3.00 \times 10^6$ (0)  &  $9.80 \times 10^0$ (5) \\
 \hline
   & $20$ & $5$  &  $0.8$  & $1.16 \times 10^3$ (5)  &  $2.62 \times 10^1$ (5) \\
INF  & $40$ & $5$  &  $0.8$  &  $9.04 \times 10^3$ (5)  & $1.86 \times 10^2$   (5) \\
    &  $60$ & $5$ &  $0.8$  &  $3.02 \times 10^4$ (5)  &  $1.71 \times 10^2$ (5)  \\
    \hline
   & $20$ & $8$  &  $0.8$  & $9.51 \times 10^3$ (5)  &   $1.70 \times 10^1$ (5) \\
INF  & $40$ & $8$  &  $0.8$  & $3.90 \times 10^5$ (5)  &  $5.20 \times 10^2$  (5) \\
    &  $60$ & $8$ &  $0.8$  & $>$ $2.01 \times 10^6$ (4) & $>$ $1.94 \times 10^3$ (0) \\

\hline
 \end{tabular}
}
}
 \label{tab:node}
\end{table}

\section{CONCLUSIONS}
\label{sec:concl}
In this paper, we formulate the non-decreasing ASFM problem into a BIP formulation with an exponential number of constraints.
For the ASFM problem, we propose an improved constraint generation algorithm that starts from a small subset of constraints and repeats solving a reduced BIP problem while adding a promising set of constraints at each iteration.
We then incorporate it into a branch-and-cut algorithm to attain good upper bounds.
According to computational results for three types of well-known benchmark instances, our algorithm performs better than the conventional modified constraint generation algorithm.



\bibliographystyle{plainnat}
\bibliography{preprint.bib}

\section*{Appendix}
The following proof is for Proposition 1. 
We prove with the following steps, (i)  $\Leftrightarrow$ (ii) $\Leftrightarrow$ (iii), (iii) $\Rightarrow$ (iv) $\Rightarrow$ (v) $\Rightarrow$ (iii).

\begin{proof} 
(i)  $\Rightarrow$ (ii).
Let $S \subseteq T \subseteq N, i \notin T, A = S \cup \{i\}, B = T$ in (i). We obtain 
\begin{equation}
f(\{i\} \mid S)  \geq   \gamma  f(\{i\} \mid T).
\end{equation}

(ii)  $\Rightarrow$ (i).
Let $\{j_1,\dots,j_l\} = A \setminus B$. We put that into (ii), and we obtain the following inequality for $i =1,\dots, l.$
\begin{equation}
f(\{j_i\} \mid A \cap B \cup \{j_1,\dots,j_{i-1}\})  \geq \gamma f(\{j_i\} \mid B \cup \{j_1,\dots,j_{i-1}\}).
\end{equation}
Sum the following $l$ equations,
\begin{equation}
\begin{array}{rll}
 f( \{j_1\} \mid A \cap B) & \geq & \gamma f( \{j_1\} \mid  B) \\
 f( \{j_2\} \mid A \cap B \cup \{j_1\}) & \geq & \gamma f( \{j_2\} \mid  B \cup \{j_1\}) \\
 & \vdots &  \\
 f(\{j_l\} \mid A \cap B \cup \{j_1,\dots, j_{l-1}\})  & \geq & \gamma \{  f(\{j_l\} \mid B \cup \{j_1,\dots,j_{l-1}\}) \} .
\end{array}
\end{equation}
We then obtain
\begin{equation}
f(A) - f(A \cap B) \geq \gamma (f(A \cup B ) -f(B)).
\end{equation}

(ii) $\Rightarrow$ (iii).
It is clear when we let $T = S \cup \{e\}$.
Since (ii) considers larger sets than (iii), 
\small
$$ \gamma = \min_{S \subseteq T \subseteq N}  \frac{f(\{i\} \mid S) }{f(\{i\} \mid T) } \leq \min_{S  \subseteq N}  \frac{f(\{i\} \mid S) } {f(\{i\} \mid S \cup \{ e\})} = \bar{\gamma}.$$ 
\normalsize

(iii) $\Rightarrow$ (ii). If $\forall S \subseteq T \subseteq N, i \notin T,~T \setminus S =\{k_1, \dots,k_q\}$, we obtain the following inequalities by (iii).
We let $S_j = S \cup \{k_1,\dots,k_j\}.$
\begin{equation}
\begin{array}{rll}
 f(\{i\} \mid S) &  \geq & \bar{\gamma} f( \{i\} \mid S \cup \{k_1\}) \\
  f(\{i\} \mid S \cup \{k_1\}) &  \geq & \bar{\gamma} f(\{i\} \mid S \cup \{k_1, k_2\})\\
  &\vdots & \\
 f(\{i\} \mid S \cup \{k_1,\dots,k_{q-1}\}) &  \geq & \bar{\gamma} f(\{i\} \mid T). 
 \end{array}
\end{equation}
By adding these inequalities, we obtain 
\begin{equation}
 f(\{i\} \mid S)  \geq  \bar{\gamma} f(\{i\} \mid T) + (\bar{\gamma} - 1) f( \{i\} \mid S \cup \{k_1\}) + \dots  + (\bar{\gamma} - 1) f(\{i\} \mid S \cup \{k_1,\dots,k_{q-1}\}).
\label{eq1}
\end{equation}
By multiplying these inequalities from the bottom, we can obtain the following relation for $f(\{i\} \mid S \cup \{k_1,\dots, k_t\})$, for $t = 1,\dots, q-1$,
\begin{equation}
 f(\{i\} \mid S \cup \{k_1,\dots, k_t\}) \geq {\bar{\gamma}}^{q-t} f(\{i\} \mid T). 
\label{eq:St}
\end{equation}
By using (\ref{eq:St}), we rewrite the inequality (\ref{eq1}) as follows.
\begin{eqnarray}
\begin{array}{rll}
 \rho_{i}(S) &  \geq & \bar{\gamma}  f(\{i\} \mid T) + (\bar{\gamma} - 1)\{ {\bar{\gamma}}^{q-1}  f(\{i\} \mid T)  + {\bar{\gamma}}^{q-2} f(\{i\} \mid T) + \dots + {\bar{\gamma}}  f(\{i\} \mid T)   \}\\
  &  = & \bar{\gamma} f(\{i\} \mid T)  + \bar{\gamma} f(\{i\} \mid T) (\bar{\gamma} - 1) \left( \frac{1-{\bar{\gamma}}^{q-1}}{1-\bar{\gamma}} \right)\\
  & = & {\bar{\gamma}}^q f(\{i\} \mid T) .
  \end{array}
\end{eqnarray}

We note that $\bar{\gamma}^q$ represents the lower bound of $\gamma$.

(iii)  $\Rightarrow$ (iv). 
For arbitrary $S$ and $T$ with $T \setminus S = \{j_1, \dots, j_l\}$ and $S \setminus T = \{k_1, \dots, k_q\}$, we obtain
\begin{equation}
\begin{array}{ll}
&  f(S \cup T) - f(S)  \\
 = & \displaystyle\sum_{t = 1}^{l}[f(S\cup\{j_1,\dots,j_t\}) - f(S \cup \{j_1,\dots, j_{t-1}\})]\\
 = &   \displaystyle\sum_{t = 1}^{l} f( \{j_t\} \mid S \cup \{j_1,\dots, j_{t-1}\})  \\
 = & f(\{j_1\} \mid S) + f(\{j_2\} \mid S \cup \{j_1\}) + f(\{j_3\} \mid S \cup \{j_1, j_2\}) + \dots  + f(\{j_l\} \mid S \cup \{j_1, \dots, j_{l-1}\})  \\
  \leq  &  f(\{j_1\} \mid S) +\frac{1}{\bar{\gamma}} f(\{j_2\} \mid S) + \frac{1}{\gamma} f(\{j_3\} \mid S ) + \dots  + \frac{1}{\gamma} f(\{j_l\} \mid S) \\
= & f(\{j_1\} \mid S) +\frac{1}{\bar{\gamma}} f(\{j_2\} \mid S) +  \displaystyle\sum_{j \in T \setminus (S \cup \{j_1, j_2\})} \frac{1}{ \gamma} f(\{j\} \mid S). 
 \end{array}
\end{equation}
Similarly, we obtain the following inequality.
\begin{equation}
\begin{array}{ll}
  & f(S \cup T) - f(T) \\
= & \displaystyle\sum_{t = 1}^{q}[f(T\cup\{k_1,\dots,k_t\}) - f(T \cup \{k_1,\dots, k_{t-1}\})]\\
= &  \displaystyle\sum_{t = 1}^{q} f(\{k_t\} \mid T \cup \{k_1,\dots, k_t\} \setminus \{k_t\} ) \\
 \geq & \displaystyle\sum_{t = 1}^{q}\gamma f(\{k_t\} \mid T \cup S \setminus \{k_t\} )  = \displaystyle\sum_{i \in S \setminus T}\gamma f(\{i\} \mid T \cup S \setminus \{i\} ). \\
 \end{array}
\end{equation}
We obtain (iv) by adding these two inequalities.

(iv) $\Rightarrow$ (v).
If $\forall S \subseteq T \subseteq N$, then $S \setminus T = \emptyset$. We obtain (v).

(v) $\Rightarrow$ (iii).
Let $ \forall S \subseteq N, T = S \cup \{j_1, j_2\}, j_1 \in N \setminus (S \cup \{j_2\})$ in (v), then we obtain
\begin{equation}
f(S \cup \{j_1, j_2\}) \leq  f(S) +  f(\{j_1\} \mid S) + \frac{1}{\bar{\gamma}} f(\{j_2\} \mid S).
\end{equation}
\begin{equation}
\begin{array}{lll}
f(\{j_2\} \mid S \cup \{j_1\}) & = & f(S \cup \{j_1, j_2\}) -f(S \cup \{j_1\}) \\
& = & f(S \cup \{j_1, j_2\}) - f(\{j_1\} \mid S) - f(S)\\
& \leq & \frac{1}{\bar{\gamma}} f(\{j_2\} \mid S).
\end{array}
\end{equation}
\end{proof}
\\

The following proof is for Proposition 2.
\begin{proof}
$(\text{i}^{\ast}) \Leftrightarrow (\text{ii}^{\ast})$. We skip the proof of $(\text{i}^{\ast})  \Rightarrow (\text{ii}^{\ast}) $ since it is the similar manner as (i)  $\Rightarrow$ (ii).

$(\text{ii}^{\ast}) \Rightarrow (\text{i}^{\ast})$.
$\gamma f(\{i\} \mid T ) \geq 0.$
Since $\gamma >0$, we obtain  $f(\{i\} \mid T ) \geq 0$ and $f(A) \leq f(B)$.
The rest is the same manner as (ii) $\Rightarrow$ (i).

$(\text{ii}^{\ast}) \Rightarrow (\text{iv}^{\ast})$. It is clear that $(\text{ii}^{\ast}) \Rightarrow$ (ii) $\Rightarrow$ (iv). 
When $f(\{i\} \mid T ) \geq 0$, the last term of (iv) is nonpositive, and that brings us $(\text{iv}^{\ast})$.

$(\text{iv}^{\ast}) \Rightarrow (\text{ii}^{\ast})$.
Suppose that $S = T \cup \{i\} $ in $(\text{iv}^{\ast})$, we obtain $f( T ) \leq f( T \cup \{i\} )$ or $ f(\{i\} \mid T ) \geq 0$ with $\gamma > 0$.
\end{proof}
\\

The following proof for Proposition 3.

\begin{proof}
($\Leftarrow$).
Suppose $\Phi \leq f(U)$, then for all $S \subseteq N$, we obtain the following inequality.
\begin{equation}
\begin{array}{l}
 f(S) +  f(\{j_1\} \mid S )y_{j_1}^{U} + \frac{1}{ \bar{\gamma}} f(\{j_2\} \mid S )y_{j_2}^{U} + \displaystyle\sum_{j \in N - (S \cup \{j_1, j_2\})} \frac{1}{ \gamma} f(\{j\} \mid S ) y_{j}^U  \\
 =  f(S) +  f(\{j_1\} \mid S )y_{j_1}^{U}  + \frac{1}{ \bar{\gamma}} f(\{j_2\} \mid S)y_{j_2}^{U}  + \displaystyle\sum_{j \in U - (S \cup \{j_1, j_2\})} {\frac{1}{ \gamma}} f(\{j\} \mid S ) y_{j}^{U} \\ 
 \geq   f(U) \geq   \Phi ,
\end{array}
\end{equation}
where the first inequality comes from Proposition 2 ($\text{iv}^{\ast}$).

%
%
%

($\Rightarrow$). If $(\Phi$, $y^U)$ $\in X$, we obtain the following inequality,
\begin{equation}
\begin{array}{lll}
\Phi &  \leq &  f(U) + f(\{j_1\} \mid U ) y_{j_1}^U  + \frac{1}{ \bar{\gamma}} \rho_{j_2}(U)y_{j_2}^U +  \displaystyle\sum_{j \in N \setminus (U \cup \{j_1, j_2\})} \frac{1}{ \gamma} \rho_j (U)y_{j}^U \\
& = & f(U).
\end{array}
\end{equation}
\end{proof}

\end{document}